# Cross comparison and modelling of Goldman Sachs, Morgan Stanley, JPMorgan Chase, Bank of America, and Franklin Resources

Ivan O. Kitov

Institute for the Geospheres Dynamics, Russian Academy of Sciences

**Abstract**
We have studied statistical characteristics of five share price time series. For each stock price, we estimated a best fit quantitative model for the monthly closing price as based on the decomposition into two defining consumer price indices selected from a large set of CPIs. It was found that there are two pairs of similar models (Bank of America/Morgan Stanley and Goldman Sachs/JPMorgan Chase) with a standalone model for Franklin Resources. From each pair, one can choose the company with the highest return depending on the future evolution of defining CPIs.

### Introduction

Four years ago we presented share price models for oil companies as based on the evolution consumer price indices (Kitov, 2009). For financial companies from the S&P 500 list, we studied the propensity to bankruptcy during the 2008/2009 period and built a number of quantitative models including those for Goldman Sachs (NYSE: GS), Morgan Stanley (MS), JPMorgan Chase (JPM), Bank of America (DAC), and Franklin Resources (BEN) (Kitov, 2010). We have been following the evolution of these five stock prices and their respective models since 2010 and found lengthy periods characterized by stable models, i.e. the models with the same defining CPIs.

Having five different models it is instructive to compare them. Our goal is to reveal similarities and differences between the models and thus between the companies. When two or more companies are driven by similar forces (same CPIs in our model) it is always helpful to understand which of the companies provides larger returns. Companies with not correlating price histories driven by different forces may be a natural choice to diversify a defensive portfolio in order to count various possible scenarios in. This article presents a feasibility study, which might be extended to a larger set of banks from and in addition to the S&P 500 list.

To begin with, we characterise five time series statistically. Cross correlation coefficients are estimated for all pairs of stock price series. Then we model all original time series and demonstrate their reliability over time. Since standard unit root tests show that these series are non-stationary, $I(1)$, processes, we (successfully) test the predicted and observed prices for cointegration. Finally, we compare the pricing models and discuss their similarity and difference in terms of investment opportunities and ideas.

### 1. Statistical estimates

Figure 1 displays the monthly closing (adjusted for splits and dividends) prices for the five studied financial companies. All curves have peaks in 2007 and troughs in 2009. There are significant differences, however. Two companies have recovered to (JPM) and above (BEN) their peak pre-crisis levels with the other three companies hovering around lower levels: 0.2 for BAC, 0.25 for MS, and 0.5 for GS, as Figure 2 shows. Table 1 lists the cross correlation coefficients for all pairs of five time series of actual monthly closing prices. All series span the interval between July 2003 and October 2012, which includes 113 readings. There are highly correlated series and not correlating ones. Not surprisingly, the cross correlation coefficient between BAC and MS, which both have been suffering most after 2007, is 0.92. At the same time, the BAC share price series does not correlate with the series from other three banks. Franklin Resources correlate with Goldman Sachs and JPMorgan Chase, with the cross correlation coefficient between the latter two companies of 0.8. Higher cross correlation coefficients suggest that driving forces behind the relevant time series are likely similar. For a quantitative model we discuss in this article, this similarity assumes close defining CPIs.

In Table 1, we also present simple statistical estimates of the model reliability, which will be discussed later on. Diagonal elements (highlighted red) are the coefficients of determination, $R^2$, as estimated from a linear regression of an actual and predicted time series for a given company. All involved series of monthly share prices are likely non-stationary processes. We have carried out several unit root tests (the augmented Dickey-Fuller and Phillips-Perron), which showed that they are all I(1) processes. (We skip technical details, which might be excessive to the broader audience. All results are available by request.) This means that cross correlation coefficients in Table 1 are subject to a positive bias.

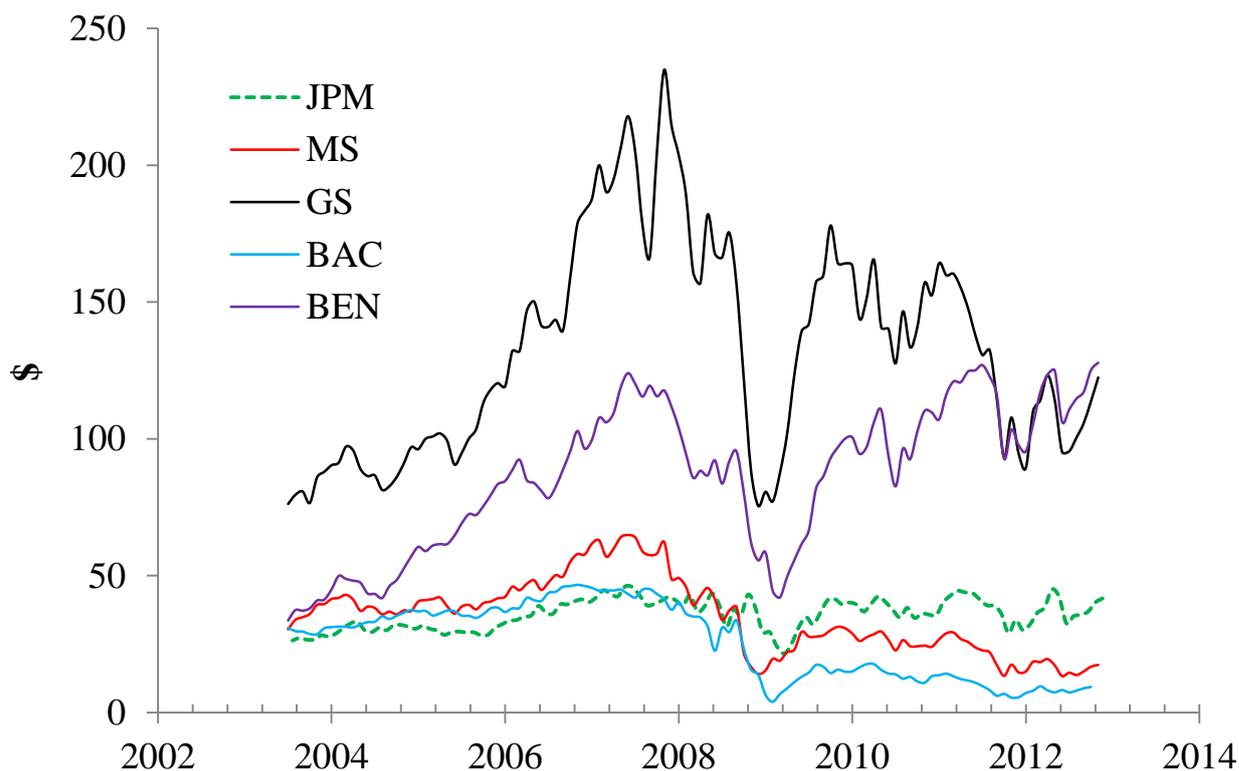

Figure 1. The evolution of JPM, MS, GS, BAC, and BEN share prices.

Table 1. Cross correlation coefficients for five time series of actual monthly closing prices. Diagonal elements (highlighted red) are the coefficients of determination, $R^2$, as estimated from a linear regression of actual and predicted time series for a given company.

|     | BAC    | BEN    | GS    | JPM   | MS    |
|-----|--------|--------|-------|-------|-------|
| BAC | 0.950  |        |       |       |       |
| BEN | -0.194 | 0.925  |       |       |       |
| GS  | 0.313  | 0.657  | 0.859 |       |       |
| JPM | 0.098  | 0.809  | 0.795 | 0.718 |       |
| MS  | 0.921  | -0.010 | 0.547 | 0.259 | 0.935 |

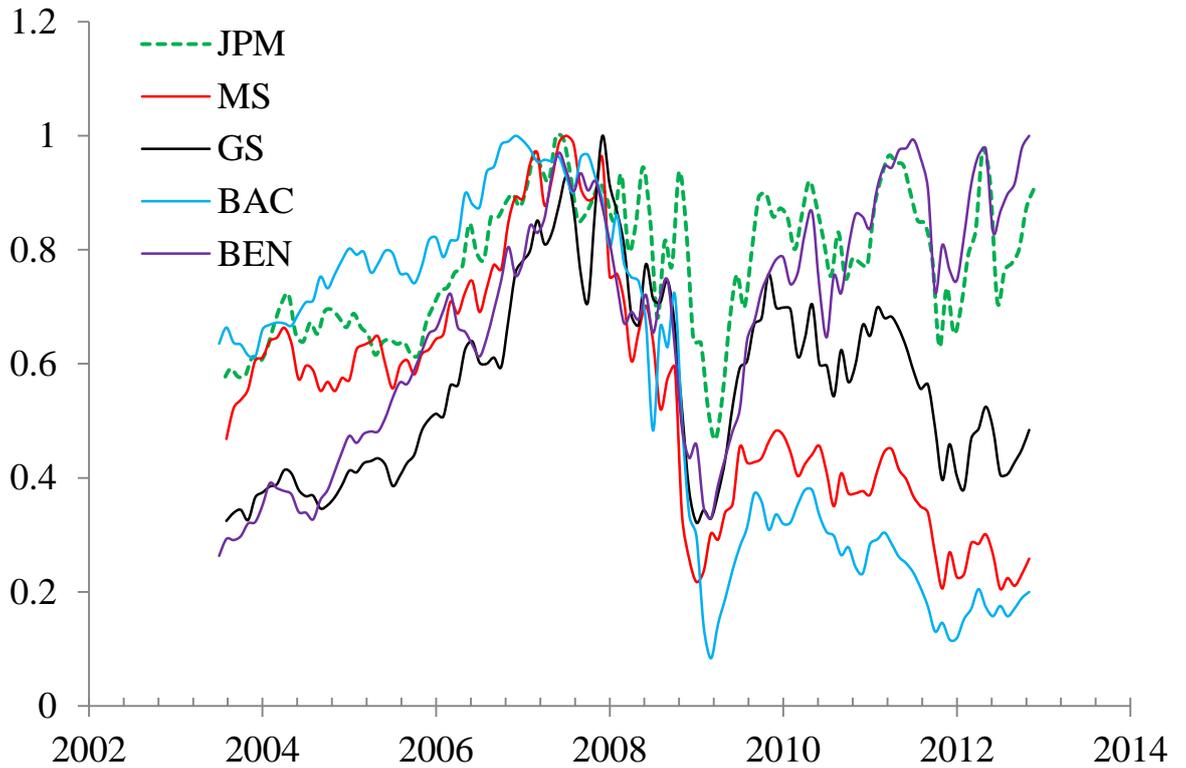

Figure 2. The evolution of JPM, MS, GS, BAC, and BEN share prices, all normalized to their peak values between 2003 and 2009.

## 1. Quantitative model

The concept of share pricing based on the link between consumer and stock prices has been under development since 2008 (Kitov, 2009). In the very beginning, we found a statistically reliable relationship between ConocoPhillips' stock price and the difference between the core and headline consumer price index (CPI) in the United States. In order to increase the accuracy and reliability of the quantitative model we extended the pool of defining CPIs to 92, which includes all major categories like food, housing, transportation etc. and many smaller subcategories. In this set, there are CPIs with similar time series, e.g. the price index of food and beverages, $F$, and the index of food only, $FB$ (Kitov, 2010). We tested the model for stability relative to these highly correlated time series.

With the extended set of defining CPIs, we estimated quantitative models for all companies from the S&P 500. A few additional companies with traded stocks were also estimated. Our model describes the evolution of a share price as a weighted sum of two individual consumer price indices selected from the set of CPIs. We allow only two defining CPIs, which may lead the modelled share price or lag behind it by several months. The intuition behind these positive and negative lags is that some companies are price setters and some are price takers. The former should influence the relevant CPIs, which include goods and services these companies produce. The latter companies lag behind the prices of goods and services they are associated with. In order to calibrate the model relative to the starting levels of the involved indices and to compensate sustainable time trends (Kitov and Kitov, 2008) (some indices are subject to secular rise or fall) we introduced a linear time trend and constant term. In its general form, the pricing model is as follows:

$$p(t_j) = \Sigma b_i \cdot CPI_i(t_j - \tau_i) + c \cdot (t_j - 2000) + d + e_j \quad (1)$$

where $p(t_j)$ is the share price at discrete (calendar) times $t_j$, $j=1,…,J$; $CPI_i(t_j-\tau_i)$ is the $i$-th component of the CPI with the time lag $\tau_i$, $i=1,...,I$ ($I=2$ in all our models); $b_i$, $c$ and $d$ are empirical coefficients of the linear and constant term; $e_j$ is the residual error, whose statistical properties have to be scrutinized. Without loss of generality, we model the monthly closing prices adjusted for splits and dividends. The monthly rate is related to the rate of CPI estimates – the frequency of output should not be larger than the frequency of input. One may use the high/low monthly prices as well as the monthly average price. We tried the monthly average of the daily closing prices and found the same models with slightly different coefficients.

By definition, the bets-fit model minimizes the RMS residual error. (One may introduce various metrics to define the best fit.) It is a fundamental feature of the model that the lags may be both negative and positive. In this study, we limit the largest lag (lead) to eleven (eight) months. System (1) contains $J$ equations for $I+2$ coefficients. We start our model in July 2003 and the share price time series has more than 100 points. To resolve the system, standard methods of matrix inversion are used.

Since November 17, 2012 we have the CPI estimates together with the monthly closing prices for October 2012. We first estimate the model with contemporary (October) readings of stock price and CPIs, with all possible CPI pairs tested with (1). Then we allow both CPIs lead (to be earlier in time) the (October) price by one and more (but less than 12) months and also estimate all possible pairs of CPI with all possible (negative) lags. For October, the best fit model has to have the smallest standard error among all estimated models.

In order to ensure that the same model was the best during a longer period before October we carry out a similar estimate for September 2012 and seven previous months. There is a big difference for these earlier models. Now one has future CPIs estimates (October, etc.) and these CPIs may lag behind the price from one (September's model) to seven (March's model) months. Thus, we have to test the models with the CPIs lagging behind the price. Otherwise, we have the same set of models as for October with all possible CPI pairs and (negative) lags from zero to eleven months. When the best fit model for September is the same as for October, i.e. defined by the same CPIs with similar lags and coefficients in (1), we consider this observation as an indication of the model reliability. For August, the defining CPIs may lag by two months and we have more models to test, both with lagging and leading CPIs. Overall, a model is considered as a reliable one when the defining CPIs are the same during eight months in a row. This number and the diversity of CPI subcategories are both crucial parameters. In further studies, it is important to extend the set of defining CPIs and the length of the model reliability. That's why quarterly revisions to all models are important. They guarantee the reliability.

Why do we rely on consumer price indices in our modelling? Many readers have reasonable doubts that some consumer price, which is not directly related to goods and services produced by a given company, may affect its price. We allow the economy to be a more complex system than described by a number of simple linear relations between share prices and goods. The connection between a firm and its products may be better expressed by goods and services which the company does not produce. The demand/supply balance is not well understood yet and may evolve along many nonlinear paths with positive and negative feedbacks. It would be too simplistic to directly define a company price by its products.

So, the intuition behind our pricing model is likely more insightful - we link a given share to some goods and services (and thus their consumer price indices), which we have to find among various CPIs. In order to provide a dynamic reference we also introduce in the model some relative and independent level of prices (also expressed by CPIs). Hence, one needs two different CPIs to define a share price model. These CPIs we select from a predetermined set of 92 CPIs by minimizing the residual model error. All in all, we assume that any share price can be represented as a weighted sum of two consumer price indices (not seasonally adjusted in our model) which may be leading the share price by several months. Our model also includes a linear time trend and an intercept in order to remove mean and trend components from all involved time series.

## 2. Modelling results

First, we report on the defining parameters for Goldman Sachs for the period between March and October 2012. Table 2 lists the best fit model for each of eight months. All models are based on the same defining CPIs – the consumer price index of food and beverages, $F$, and the index of owners' rent of primary residence, $ORPR$. In all cases, the lags are the same: three and two months, respectively. Other coefficients and the standard error suffer just slight oscillations or drifts (e.g. $c$ and $d$). It is important to stress again that all models the months except October also include those with the future CPIs. Table 2 confirms that no future CPIs drive the share price since March 2012. This company may be considered as a price setter. For the purposes of this study, we use the following best fit model for GS:

$$GS(t) = -13.795F(t-3) + 11.027ORPR(t-2) + 29.935(t-2000) + 33.751, \text{sterr}=\$14.52 \quad (2)$$

Table 2. The monthly models for GS.

| Month | $C_1$ | $t_1$ | $b_1$ | $C_2$ | $t_2$ | $b_2$ | $c$ | $d$ | sterr,$ |
|---|---|---|---|---|---|---|---|---|---|
| October | F | 3 | -13.795 | ORPR | 2 | 11.027 | 29.935 | 33.751 | 14.521 |
| September | F | 3 | -13.791 | ORPR | 2 | 11.013 | 29.992 | 35.827 | 14.584 |
| August | F | 3 | -13.787 | ORPR | 2 | 11.003 | 30.023 | 37.106 | 14.649 |
| July | F | 3 | -13.759 | ORPR | 2 | 10.978 | 30.018 | 37.647 | 14.707 |
| June | F | 3 | -13.731 | ORPR | 2 | 10.933 | 30.124 | 41.985 | 14.758 |
| May | F | 3 | -13.704 | ORPR | 2 | 10.876 | 30.342 | 48.755 | 14.770 |
| April | F | 3 | -13.661 | ORPR | 2 | 10.819 | 30.449 | 53.171 | 14.805 |
| March | F | 3 | -13.787 | ORPR | 2 | 10.943 | 30.440 | 48.639 | 15.055 |

In Tables 3 through 6, we summarize the evolution of models for four banks since March 2012. Taking the defining CPIs and coefficients for October 2012 one obtains the following best fit models:

$$BAC(t) = -5.897SEFV(t-0) + 2.650RSH(t-2) + 20.609(t-2000) + 444.030, \text{sterr}=\$2.98 \quad (3)$$
$$MS(t) = -7.93SEFV(t-0) + 4.415ORPR(t-2) + 25.226(t-2000) + 420.919, \text{sterr}=\$3.47 \quad (4)$$
$$JPM(t) = -1.856F(t-4) + 0.993ORPR(t-2) + 7.037(t-2000) + 116.907, \text{sterr}=\$2.96 \quad (5)$$
$$BEN(t) = -7.333FB(t-4) - 1.519O(t-9) + 69.578(t-2000) + 1536.224, \text{sterr}=\$7.36 \quad (6)$$

where $SEFV$ is the consumer price index of food away from home, $RSH$ is the index of rent of shelter, $FB$ is the index of food without beverages, and $O$ is the index of other goods and services. Therefore, all five models include the index related to food. (We consider April's JPM model as a fluctuation.) Figure 3 shows that $FB$ and $F$ are practically identical and we might exclude one of them from the full set of CPIs without any significant loss in resolution. On the other hand, the BEN model is stable with $FB$ and we retain it in the set.

In four from five models, the second CPI is associated with rent of residence ($ORPR$) or shelter ($RSH$). Figure 3 demonstrates that these indices are also close. Table 7 lists cross correlation coefficients, $CC$, for the six defining CPIs and their first differences. Because of secular growth in prices, these coefficients are extremely high for the original series, but these levels are likely biased up. The first differences characterize the link between the indices in a more reliable way, with $CC=0.994$ for the first differences of $F$ and $FB$. The first difference of $SEFV$, $dSEFV$, is well correlated with $dF$ and $dFB$. Taking into account all possible time lags between the indices (from 0 to 11 months) in the models one may calculate cross correlation coefficients for the same time series but with various time shifts. Obviously, the highest cross correlation coefficients should not be lower than that for the contemporary time series. In Table 7, the highest *CC*s among all time lags are shown in brackets. For example, the *CC* for d*SEFV* and d*F/dFB* has increased to 0.49. Interestingly, the first difference of *SEFV*, d*SEFV*, has the same correlation coefficient with d*ORPR* as d*RSH*, but d*SEFV* and d*RSH* do not correlate. When time lags between the indices are allowed, no big change in the level of correlation of d*SEFV* and d*ORPR* is observed. Overall, it is possible to distinguish three different sets of CPIs: "food", "rent", and "other".

Figure 4 depicts all five models as compared to the relevant actual prices since July 2003. We also plotted the high/low monthly pricing in order to illustrate the level of fluctuations of the intermonth prices. One may model the monthly closing prices as well as the high, low, average, etc. prices and likely obtain slightly different models. As mentioned above, we have estimated $R^2$ for five models, as Table 1 lists. All coefficients of determination are larger than 0.7, with three from five models having $R^2>0.9$. In order to prove that these statistical estimates for our quantitative models are not biased we have tested them for cointegration between actual and predicted series. The Johansen tests for cointegration rank have shown cointegration rank 1 in all cases. We have also tested the model residual time series (see Figure 5) for unit roots and found that they are I(0) processes. Therefore the predicted and observed series are cointegrated for all banks and no $R^2$ in Table 1 is biased.

Table 3. The monthly models for BAC. The last column lists standard errors.

| Month | $C_1$ | $t_1$ | $b_1$ | $C_2$ | $t_2$ | $b_2$ | $c$ | $d$ | sterr,$ |
|---|---|---|---|---|---|---|---|---|---|
| October | SEFV | 0 | -5.897 | RSH | 2 | 2.650 | 20.609 | 444.030 | 2.983 |
| September | SEFV | 0 | -5.906 | RSH | 2 | 2.656 | 20.625 | 444.228 | 2.979 |
| August | SEFV | 0 | -5.965 | RSH | 2 | 2.679 | 20.868 | 448.932 | 2.962 |
| July | SEFV | 0 | -5.953 | RSH | 2 | 2.684 | 20.751 | 446.137 | 2.953 |
| June | SEFV | 0 | -5.989 | RSH | 2 | 2.695 | 20.924 | 449.647 | 2.952 |
| May | SEFV | 0 | -5.982 | RSH | 2 | 2.699 | 20.850 | 447.823 | 2.949 |
| April | SEFV | 0 | -5.960 | RSH | 2 | 2.690 | 20.757 | 446.303 | 2.949 |
| March | SEFV | 0 | -5.971 | RSH | 2 | 2.698 | 20.772 | 446.266 | 2.947 |

Table 4. The monthly models for MS.

| Month | $C_1$ | $t_1$ | $b_1$ | $C_2$ | $t_2$ | $b_2$ | $c$ | $d$ | sterr,$ |
|---|---|---|---|---|---|---|---|---|---|
| October | SEFV | 0 | -7.93 | ORPR | 2 | 4.415 | 25.226 | 420.919 | 3.468 |
| September | SEFV | 0 | -7.90 | ORPR | 2 | 4.399 | 25.137 | 420.060 | 3.468 |
| August | SEFV | 0 | -7.96 | ORPR | 2 | 4.425 | 25.343 | 423.817 | 3.447 |
| July | SEFV | 0 | -7.96 | ORPR | 2 | 4.445 | 25.258 | 420.687 | 3.440 |
| June | SEFV | 0 | -8.01 | ORPR | 2 | 4.449 | 25.526 | 426.655 | 3.437 |
| May | SEFV | 0 | -8.01 | ORPR | 2 | 4.452 | 25.540 | 426.579 | 3.434 |
| April | SEFV | 0 | -7.97 | ORPR | 2 | 4.419 | 25.492 | 427.246 | 3.422 |
| March | SEFV | 0 | -8.00 | ORPR | 2 | 4.431 | 25.609 | 429.254 | 3.421 |

Table 5. The monthly models for JPM.

| Month | $C_1$ | $t_1$ | $b_1$ | $C_2$ | $t_2$ | $b_2$ | $c$ | $d$ | sterr,$ |
|---|---|---|---|---|---|---|---|---|---|
| October | F | 4 | -1.856 | ORPR | 2 | 0.993 | 7.037 | 116.907 | 2.955 |
| September | F | 4 | -1.859 | ORPR | 2 | 1.006 | 6.965 | 114.846 | 2.932 |

| Month | | | | | | | | | |
|---|---|---|---|---|---|---|---|---|---|
| August | F | 4 | -1.861 | ORPR | 2 | 1.018 | 6.898 | 112.917 | 2.914 |
| July | F | 4 | -1.863 | ORPR | 2 | 1.024 | 6.873 | 112.112 | 2.912 |
| June | F | 4 | -1.865 | ORPR | 2 | 1.024 | 6.883 | 112.342 | 2.912 |
| May | F | 4 | -1.863 | ORPR | 2 | 1.024 | 6.877 | 112.182 | 2.912 |
| April | FH | 4 | -1.254 | FAB | 2 | 1.770 | 10.219 | -12.460 | 2.905 |
| March | F | 4 | -1.878 | ORPR | 2 | 1.051 | 6.791 | 109.260 | 2.839 |

Table 6. The monthly models for BEN.

| Month | $C_1$ | $t_1$ | $b_1$ | $C_2$ | $t_2$ | $b_2$ | c | d | sterr,$ |
|---|---|---|---|---|---|---|---|---|---|
| October | FB | 4 | -7.333 | O | 9 | -1.519 | 69.578 | 1536.224 | 7.365 |
| September | FB | 4 | -7.319 | O | 9 | -1.515 | 69.428 | 1533.079 | 7.361 |
| August | FB | 4 | -7.301 | O | 9 | -1.513 | 69.275 | 1529.960 | 7.353 |
| July | FB | 4 | -7.299 | O | 9 | -1.515 | 69.286 | 1530.175 | 7.353 |
| June | FB | 4 | -7.311 | O | 9 | -1.515 | 69.375 | 1532.037 | 7.350 |
| May | FB | 4 | -7.301 | O | 9 | -1.513 | 69.304 | 1529.705 | 7.343 |
| April | FB | 4 | -7.303 | O | 9 | -1.515 | 69.361 | 1530.410 | 7.337 |
| March | FB | 4 | -7.309 | O | 9 | -1.513 | 69.312 | 1531.360 | 7.270 |

Table 7. Cross correlation coefficients for six CPI time series and their first differences. Original series include 124 readings, and their first differences – 123 readings.

| | F | FB | SEFV | ORPR | RSH | O |
|---|---|---|---|---|---|---|
| F | 1 | | | | | |
| FB | 0.99998 | 1 | | | | |
| SEFV | 0.99714 | 0.99671 | 1 | | | |
| ORPR | 0.98356 | 0.98295 | 0.98702 | 1 | | |
| RSH | 0.97533 | 0.97478 | 0.97736 | 0.99698 | 1 | |
| O | 0.97752 | 0.97661 | 0.98664 | 0.95629 | 0.93924 | 1 |

| | dF | dFB | dSEFV | dORPR | dRSH | dO |
|---|---|---|---|---|---|---|
| dF | 1 | | | | | |
| dFB | 0.994 | 1 | | | | |
| dSEFV | 0.47 [0.49] | 0.48 [0.49] | 1 | | | |
| dORPR | 0.12 [0.26] | 0.12 [0.26] | 0.31 [0.35] | 1 | | |
| dRSH | 0.13 [0.30] | 0.12 [0.28] | 0.10 [0.29] | 0.31 [0.37] | 1 | |
| dO | -0.18 [030] | -0.18 [0.28] | 0.06 [0.29] | 0.002 [-0.21] | 0.04 [-0.29] | 1 |

**Discussion**

The stock prices of BAC and MS are well correlated. This observation is supported by the similarity of defining CPIs with equal time lags. It is worth noting that the level of correlation may cease quickly for two models with the same defining CPIs but with increasing difference in time lags. For the same CPIs and lags, the level of correlation depends on the ratio of CPI coefficients. This ratio ($b_1/b_2$) is -2.23 for BAC and -1.65 for MS. The closeness of the ratios guarantees similar evolution of their prices. It is important to stress, however, that *SEFV* has a relatively lower influence on BAC price than *RSH*. The ratio of the current level of share price to $b_1$ defines the sensitivity of the share prices to *SEFV*. For BAC, this ratio is -1.52, and for MS is -2.1. In other words, one unit change in *SEFV* forces a $5.9 (~65%) fall in BAC and a $8 or 47% fall in MS. Depending on the future absolute evolution of *SEVF* and its evolution relative to *RSH* (*ORPR* for MS) one may quantitatively estimate the performance of BAC and MS and choose the company to invest in. If food price rises at a higher rate than that of rent of residence one may prefer Morgan Stanley.

Goldman Sachs and JPMorgan Chase have the same defining CPIs (*F* and *ORPR*) and practically the same time lags. The ratio of coefficients is -1.23 and -1.87, respectively. According to the ratio of share price to $b_1$, JPM is less sensitive to food prices, i.e. one unit change in *F* forces a fall by about $14 in GS (~11%) and only $2 or 4% in JPM. Therefore, GS stock price fell lower from its peak in 2007 and recovered in 2009 to the 0.8 of the pre-crisis level due to a deep and quick fall in food prices. The fall in GS price in 2011 was induced by a surge in food prices. In 2012, d*F* was slightly lower than d*ORPR* and both were positive. As a result, both prices have been oscillating around constant levels. When the food price falls, one should choose GS. In the case of food price growth, JPM looks better.

The BEN model contains a different defining CPI (*O*) which has a quite specific shape with a high-amplitude step between February and April 2009. Instructively, the first difference of *O* does not correlate with any other involved index. For BEN, the step in *O* series is associated with a sharp fall in the stock price nine months before, as the negative coefficient in Table 6 assumes. One may suggest that despite all companies had the same fall around the same time their further evolution resulted in different models. We interpret this observation as an indication that BEN stocks are driven by some forces different from other companies. Despite its high correlation with MS, the price of BEN may deviate much in the future and corrupt the correlation. For BEN, the best situation is when the defining prices do not grow fast.

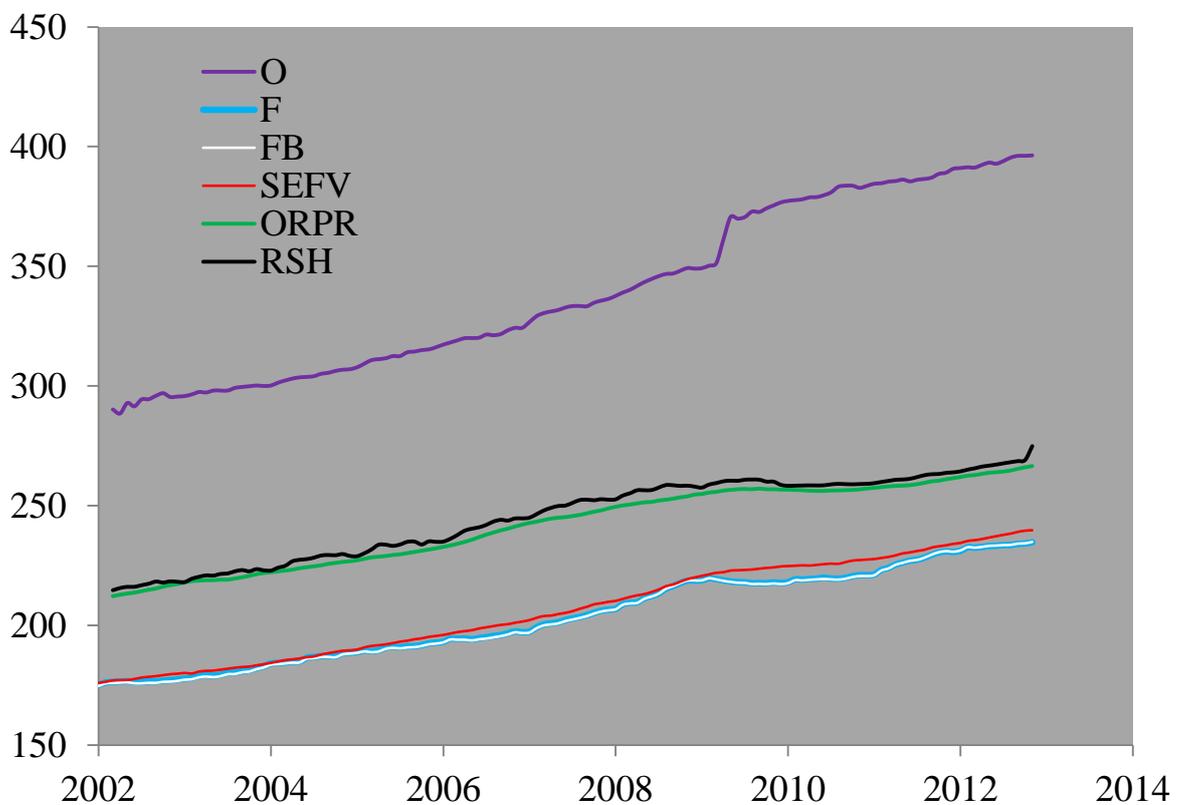

Figure 3. The evolution of all defining CPIs. Notice F (blue line) and FB (white line inside the blue line) are practically identical

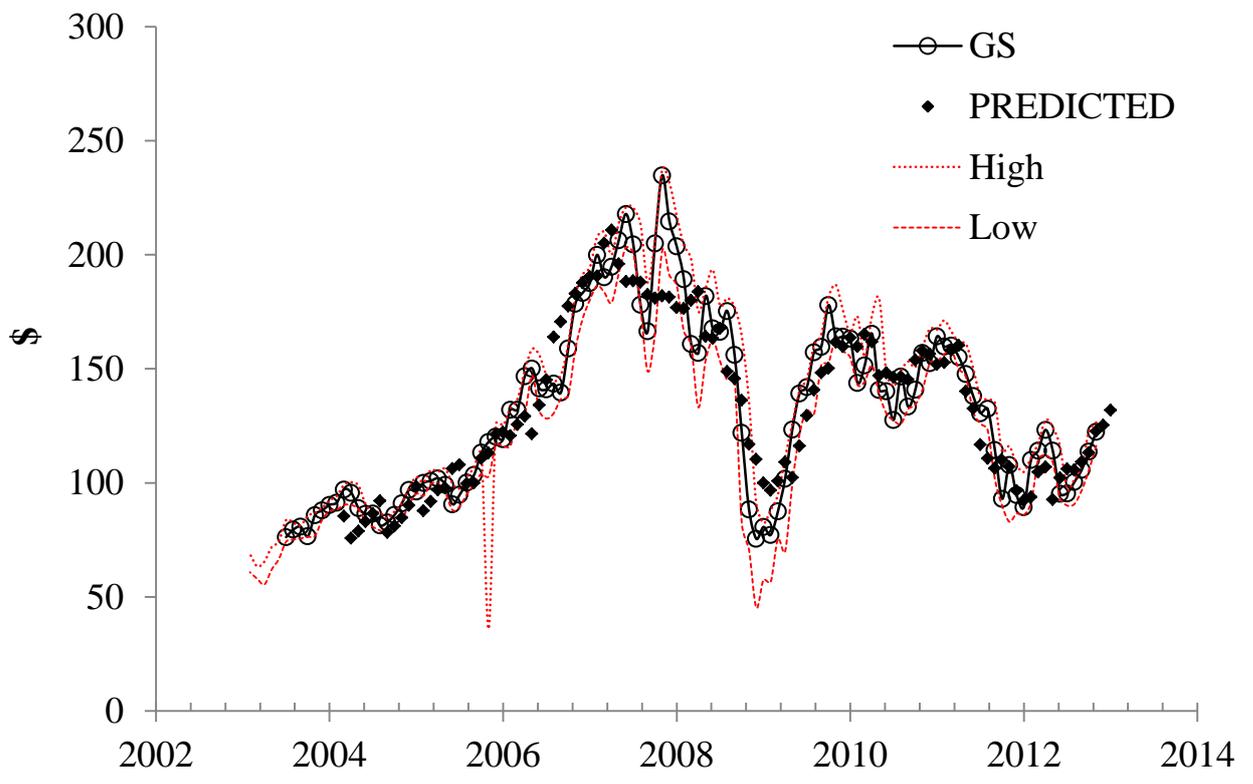

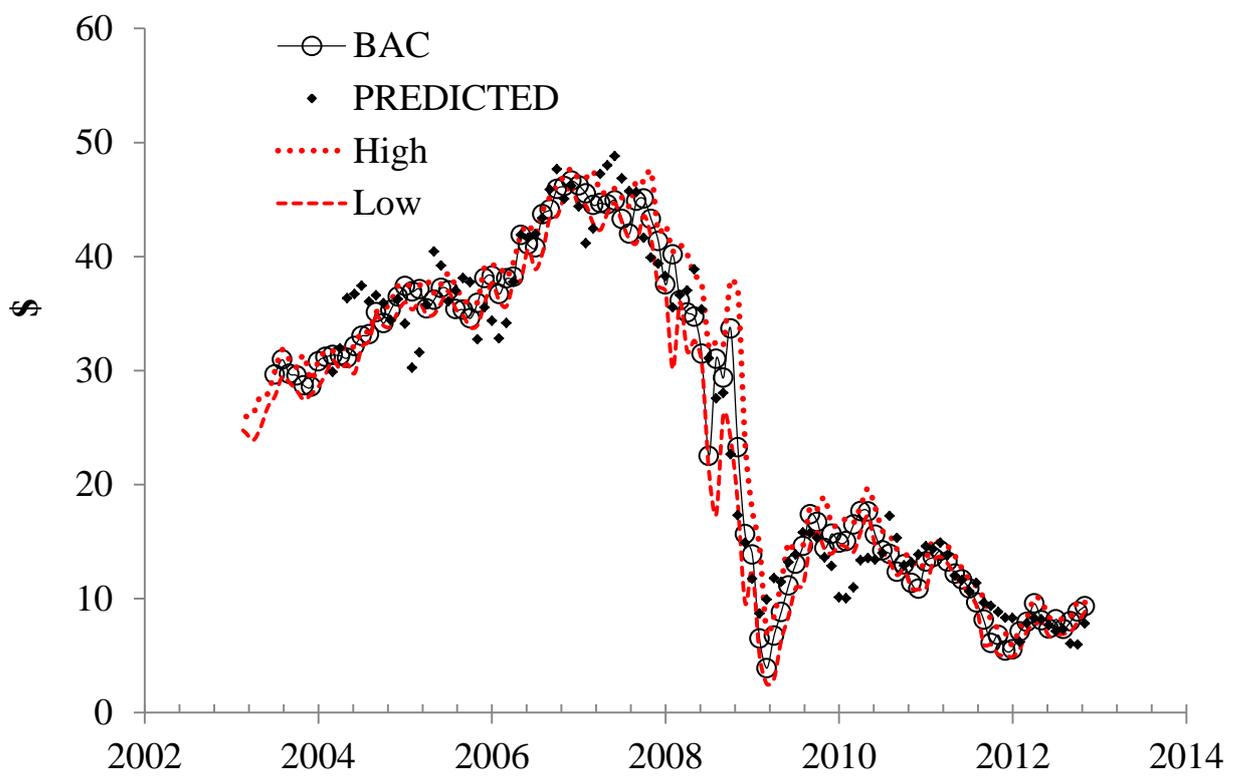

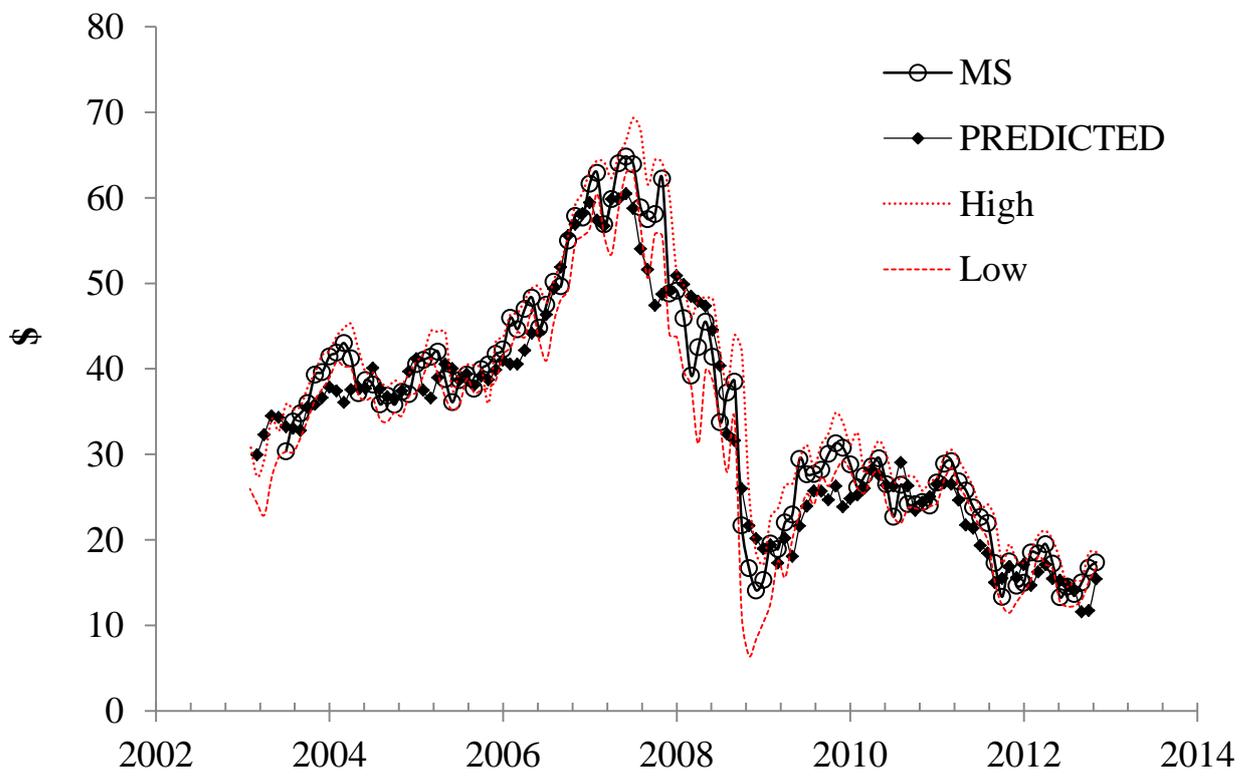

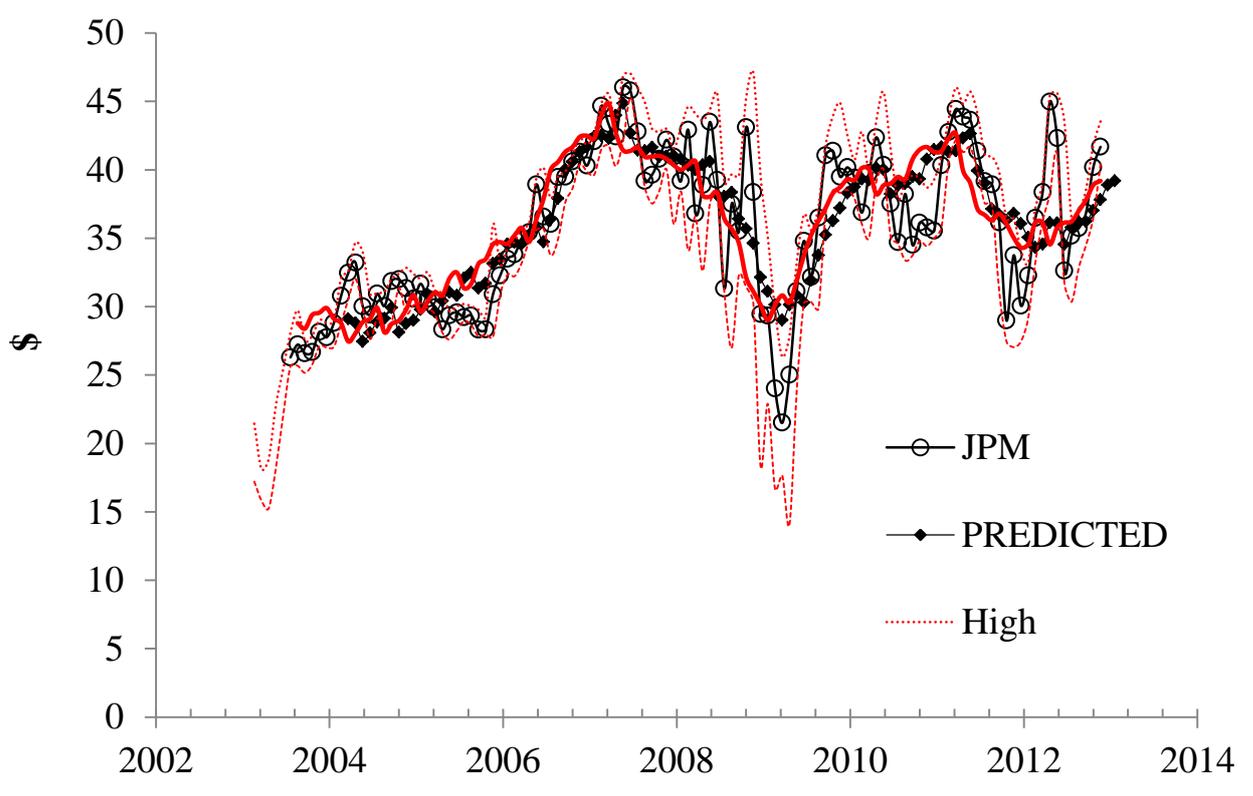

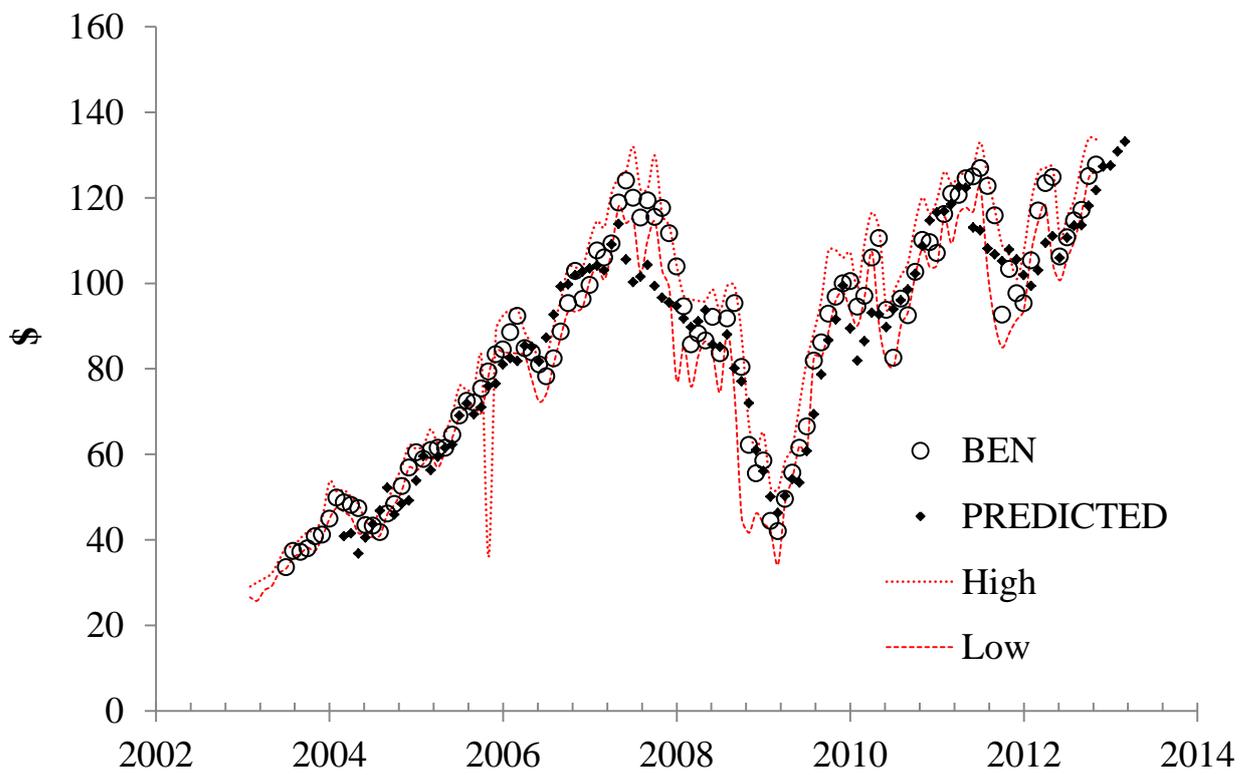

Figure 4. Observed and predicted share prices together with their high/low monthly prices.

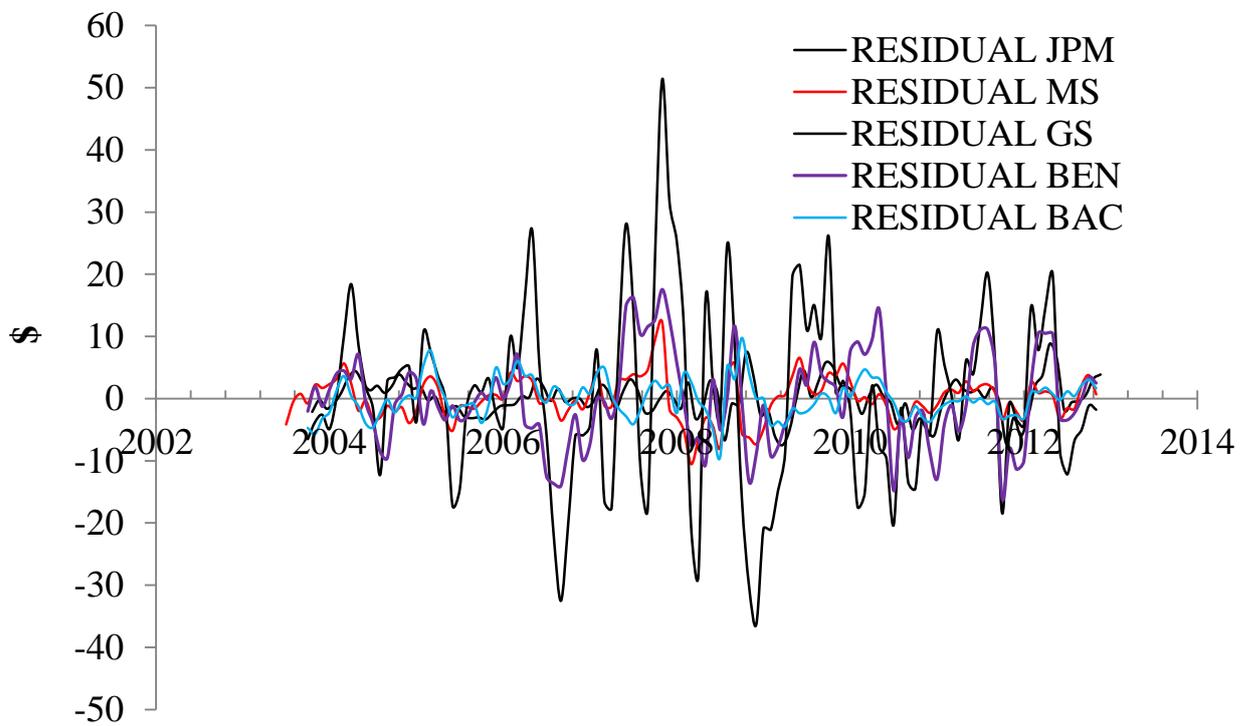

Figure 5. The residual model errors.

# References

Kitov, I. (2009). Predicting ConocoPhillips and Exxon Mobil stock price, Journal of Applied Research in Finance, Spiru Haret University, Faculty of Financial Management and Accounting Craiova, vol. I(2(2)_ Wint), pp. 129-134.

Kitov, I. (2010). Modelling Share Prices of Banks and Bankrupts, Theoretical and Practical Research in Economic Fields, Association for Sustainable Education, Research and Science, vol. 0(1), pages 59-85, June.

Kitov, I. and O. Kitov (2008). Long-Term Linear Trends In Consumer Price Indices, Journal of Applied Economic Sciences, Spiru Haret University, Faculty of Financial Management and Accounting Craiova, vol. 3(2(4)_Summ).